\documentstyle[12pt,epsfig]{article}
\let\section=\subsection     \let\subsection=\subsubsection                
\newcommand{\beq}{\begin{equation}}
\newcommand{\eeq}{\end{equation}}
\newcommand{\bea}{\begin{eqnarray}}
\newcommand{\eea}{\end{eqnarray}}
\newcommand{\bfg}{\begin{figure}}
\newcommand{\efg}{\end{figure}}
\newcommand{\ie}{{\it i.e.}}
\newcommand{\eg}{{\it e.g.}}
\newcommand{\etal}{{\it et al.}}

\begin{document}
\begin{center}
{\large \bf ANTI-BARYON PUZZLE IN ULTRA-RELATIVISTIC HEAVY-ION COLLISIONS}
\\[2mm]
   R.~RAPP and E.V.~SHURYAK\\[5mm]
   {\small \it  Department of Physics and Astronomy, \\
   SUNY Stony Brook, New York 11794-3800, USA \\[8mm] }
\end{center}

\begin{abstract}\noindent
The evolution of (non-strange) antibaryon abundances 
in the hadronic phase of central heavy-ion collisions is studied 
within a thermal equilibrium framework, based on the well-established 
picture of subsequent chemical and thermal freezeout. Due to large 
annihilation cross sections, antiprotons are, a priori, not expected 
to comply with this scheme. However, we show that a significant 
regeneration of their abundance occurs upon inclusion of the inverse
reaction of multipion fusion, $n_\pi\pi \to p\bar p$ 
(with $n_\pi$=5-6), necessary to ensure detailed balance.
Especially at SPS energies, the build-up of large pion-chemical
potentials between chemical and thermal freezeout reinforces this 
mechanism, rendering the $\bar p/p$ ratio 
in reasonable agreement with the observed one (reflecting 
chemical freezeout).    
Explicit solutions of the pertinent rate equation, which account
for chemical off-equilibrium effects, corroborate this explanation. 
\end{abstract}

\section{Introduction}   
The combined experimental and theoretical study of (ultra-) 
relativistic heavy-ion 
collisions has lead to impressive progress in our understanding
of the mechanisms underlying the formation and evolution of 
strongly interacting matter. A key issue concerns the question 
whether one is indeed producing systems of sufficient macroscopic
extent with frequent enough re-interactions among the constituents 
that justify the use of (thermal-) equilibrium techniques. This, 
after all, constitutes an essential prerequisite for a meaningful 
investigation of the phase diagram of QCD in these reactions, 
including the identification of (the approach towards) phase 
transitions and Quark-Gluon Plasma (QGP) formation.   
A successful interpretation of experimental observables has to
be based on a comprehensive overall picture of the collision
dynamics. For SPS energies, this has been accomplished to a
remarkable degree, as evidenced by, \eg, 
(i) hadro-chemical equilibration as inferred from produced
particle ratios, 
(ii) thermal equilibration with collective transverse 
expansion, as inferred from transverse momentum spectra  
combined with two-particle correlation measurements,
(iii) a substantial excess of electromagnetic radiation (photons and
dileptons) ascribed to thermal emission, 
(iv) the 'anomalous' suppression pattern of charmonium states. 
An important feature of this picture is the distinction between
{\em chemical} and {\em thermal} freezeout, implying a significant 
duration of an expanding hadronic phase (also essential to explain
(ii) and (iii)).  

A solid understanding of the majority of observables then allows to 
go beyond and investigate well-defined deviations. 
In this talk we will focus on antiproton production, which, upon closer
inspection, is not easily understood within the above picture,
but, as we argue below, can be reconciled with it.       

The paper is organized as follows:
in Sect.~\ref{sec_equil} we briefly recall the essential elements 
of hadronic equilibration needed to formulate the 
'antiproton puzzle'. In Sect.~\ref{sec_rateq} we address the problem 
within a thermal rate equation, first for the more transparent 
equilibrium case, then including off-equilibrium effects via 
its explicit solution. We summarize in Sect.~\ref{sec_sum}.

\section{Equilibration Times and Antiproton Puzzle}
\label{sec_equil}
\subsection{Two Types of Hadronic Freezeout}
The notion of chemical and thermal freezeout stages  
in central collisions of heavy nuclei can be given a 
clear physical meaning (and applies equally well to
hadronic and partonic degrees of freedom).
It is based on a hierarchy of time scales related to the 
underlying cross sections. 

{\em Thermal} equilibrium in the hadronic fireball can be (locally) 
maintained as long as {\em elastic} collisions are able to keep up 
with its expansion (dilution) rate. This criterium can be quantified 
as
\beq 
R(\tau) \equiv 
\int \limits_\tau^\infty \frac{d\tau'}{\tau_{th}(\tau')} 
\le R_{fo} 
\label{Rtau}
\eeq 
with $R_{fo}\simeq 1/2$ (almost independent of collision energy from
SIS/BEVALAC to SPS) as extracted from hydrodynamic fits to 
single-particle spectra~\cite{MS83,HS98}. The thermal relaxation
time for a hadron species $i$ is given by 
\beq
\left( \tau_{th,i} \right)^{-1}= \sum\limits_h \langle \sigma_{ih}
 v_{rel} \rangle \ \rho_h  
\eeq 
in terms of the density $\rho_h$ of hadrons which serve as scattering 
centers, and the pertinent (thermally averaged) cross section
$\langle \sigma_{ih} \rangle$ ($v_{rel}$: relative velocity between
hadron $i$ and $h$). 
Standard (resonance dominated) hadronic reactions, such as 
$\pi\pi\to \rho \to \pi\pi$, 
$\pi K \to K^* \to \pi K$, $\pi N \to \Delta \to \pi N$ (or even 
nonresonant $NN$ scattering), lead to typical values around   
$\langle \sigma\rangle \simeq 50$~mb. Even for a rather dilute hadron 
gas of normal nuclear matter density ($\rho_h=\rho_0=0.16$~fm$^{-3}$)
one finds a thermal equilibration time as short as 2~fm/c.    
A simplified freezeout criterium, which does not explicitly involve 
the expansion time scale,  is given by the condition that 
the (thermally averaged) mean free path becomes equal to the system size,
$\lambda_{MFP}\simeq r_{FB}$ ($r_{FB}$: fireball radius).  
Both criteria, when applied to central heavy-ion collisions at full
SPS energy ($E_{lab}=158$~AGeV), indicate thermal freezeout temperatures
of about $T_{th}=110-120$~MeV.    
This is consistent with combined experimental analysis of (light-) hadron 
$p_t$-spectra and HBT correlations~\cite{Stock99}, which can be 
schematically summarized by (nonrelativistic) slope parameters 
$T_{slope,i}=T_{th}+m_i v_t^2$
with an average transverse flow velocity $v_t\simeq$~0.4-0.5~$c$ at 
thermal freezeout.  
 
{\em Chemical} equilibrium is maintained by {\em inelastic} collisions 
which change the particle composition of the system. 
Corresponding hadronic cross sections, \eg,  for 
$\pi\pi \leftrightarrow K\bar K,\rho\rho$, are typically 
much smaller than elastic ones, 
$\langle \sigma_{inel} \rangle \simeq 1$~mb after thermal averaging. 
Consequently, chemical equilibration times are around 
$\tau_{chem} \simeq 100$~fm/c, well above possible lifetimes of 
hadronic fireballs created in high-energy nuclear collisions.  
If anything, the chemical composition of the hadronic phase must frozen 
early in the evolution. At full SPS energy essentially all hadron 
abundances can be accommodated by a common chemical freezeout at 
$(T_{ch},\mu_B^{ch})\simeq (170,260)$~MeV~\cite{pbm99,beca00}. 
This result is, in fact, 
difficult to understand in purely hadronic scenarios and has been 
interpreted as 'prehadronic' flavor equilibration, possibly in a QGP.

A particular consequence of the sequential freezeouts that will be 
important below is the effective 
conservation of particle numbers for species 
that are not subject to strong decays (\eg, $\pi$, $K$, $\eta$;  
after all, the abundances at thermal
freezeout have to agree with the measured ones).
In statistical mechanics language this is expressed by finite 
meson-chemical potentials, which, under SPS conditions, reach  
appreciable values of $\mu_\pi^{th}\simeq$~60-80~MeV, 
$\mu_K^{th} \simeq$~100-130~MeV (similar for $\mu_\eta^{th}$), etc.
Resonances that are regenerated via strong interactions  
acquire the chemical potentials of their decay products, 
\ie, $\mu_\Delta=\mu_N+\mu_\pi$, $\mu_\rho=2\mu_\pi$, 
$\mu_{K^*}=\mu_K+\mu_\pi$, etc.\footnote{Strongly decaying resonances 
with narrow widths ($\le$~50~MeV or so) are in between the two cases. 
However, for the vector mesons $\omega$ and $\phi$, \eg, it turns out
that the two limiting cases  -- full chemical equilibrium 
($\mu_\omega=3\mu_\pi$ and $\mu_\phi=2\mu_k$) versus chemical 
off-equilibrium (individual $\mu_\omega$ and $\mu_\phi$ adjusted to their
conserved number) -- give numerically very similar results. We 
also note that, if reactions of the type $\omega\pi\leftrightarrow \pi\pi$
are relevant, the $\omega$ chemical potential would be reduced.}   

\bfg[!ht]
\begin{minipage}[t]{6.8cm}
\vspace{-0.1cm}
\epsfig{file= muTrhic-sps2.eps,width=6.7cm}
\vspace{-0.2cm}
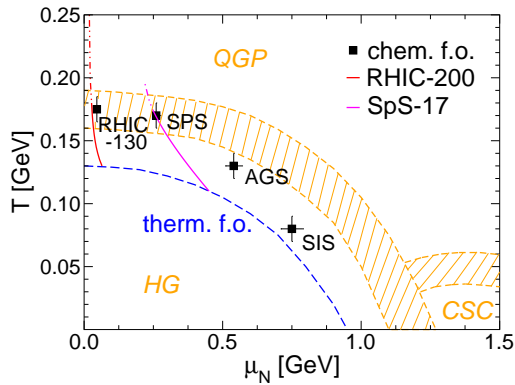
\caption{Schematic QCD phase diagram including empirical information
on chemical freezeout in (ultra-) relativistic heavy-ion collisions.}
\vspace{-0.5cm}
\label{fig_phasedia}
\end{minipage}\hfill
\begin{minipage}[t]{6.7cm}
\vspace{-0.2cm}
\epsfig{file=mu_N_pi-SpS.eps,width=6.7cm}
\vspace{-0.8cm}
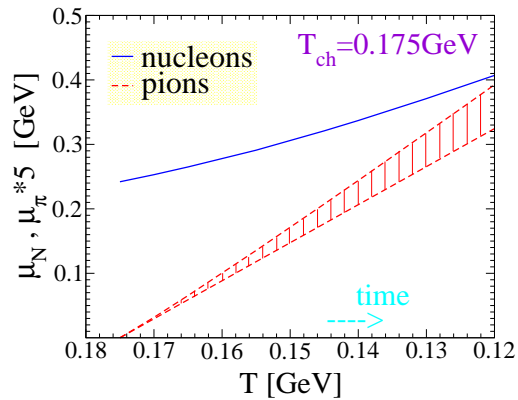
\caption{Evolution of nucleon- and 
pion-chemical potentials along the SPS 
trajectory of Fig.~\protect\ref{fig_phasedia}, including   
un\-cer\-tain\-ties in $\mu_\pi^{therm}\simeq$~65-80~MeV.}
\vspace{-0.5cm}
\label{fig_chemevo}
\end{minipage}
\efg

\subsection{Antiproton Systematics}
An enhanced production of antiprotons in central heavy-ion collisions
(over the extrapolation from $pp$ or $pA$ reactions) has been among
suggested signatures for QGP formation~\cite{HSSG86,KMSG88} (which is
 expected to facilitate copious antiquark production). The experimental 
results for $\bar p/p$ ratios at AGS~\cite{e917}, SPS~\cite{na44,na49} 
and RHIC~\cite{star,phenix} are, however, consistent with the chemical 
freezeout systematics at the respective energies~\cite{pbm99,beca00,pbm95}.
In Pb(158~AGeV)+Pb, one finds~\cite{na44} $\bar p/p=(5.5\pm 1)$~\%, 
to be compared to $\bar p/p = \exp(-2\mu_N^{ch}/T_{ch}) \simeq 5$\%. 
Moreover, the $\bar p/p$ ratio does not exhibit a pronounced centrality 
dependence~\cite{na49}, 
which is also borne out of hadro-chemical model fits~\cite{CKW02}.  

Despite this apparent agreement it is nevertheless puzzling why the number 
of antiprotons should to be 'frozen in' early in the hadronic phase: 
large $\bar pp$ annihilation cross section of around 50~mb at the relevant
thermal energies ($E_p^{th}=m_p+\frac{3}{2} kT$), together with the 
sizable baryon densities under SPS conditions, seem to imply a chemical 
freezeout for antibaryons at a much lower temperature (\ie, later time) 
than $T_{ch}\simeq 170$~MeV; \eg, for $T=150$~MeV, where according to 
thermal fireball calculations~\cite{RW99} $\rho_B\simeq 0.75~\rho_0$, 
one obtains for the chemical relaxation time, 
\beq
\tau_{\bar p}^{chem}
=\frac{1}{\langle \sigma_{p\bar p}^{ann}(s) v_{rel} \rangle \ \rho_B} \ , 
\label{tauchem}
\eeq
a value of about 3~fm/c, 
which is significantly smaller than the fireball lifetime.  
A naive evaluation of the $\bar p/p$ ratio using thermal freezeout   
conditions, $(T_{th},\mu_N^{th})\simeq (120,400)$~MeV~\cite{RW99}, 
yields $\exp(-2\mu_N^{th}/T_{th})\simeq 0.1$~\%,
a factor $\sim$~50 below the measured value! This constitutes 
the 'antiproton puzzle'. 

There have been attempts to resolve this puzzle within sequential 
scattering (transport) simulations~\cite{arc97,urqmd00}. 
The latter are, in principle, suited for a realistic description 
of the late (dilute) hadronic
stages in heavy-ion collisions. Nevertheless, extra assumptions, 
such as a strong in-medium suppression ('shielding') of the annihilation
cross section~\cite{arc97} or an enhanced production via an increased 
string tension in the early phases~\cite{urqmd00}, had to be invoked to
avoid or compensate annihilation losses in the hadron gas. 
However, a notorious problem within these approaches is the treatment
of multi-body collisions (3 or more), which are usually neglected, 
thus violating detailed balance. For the case at hand, these are 
multi-meson fusion reactions, \eg, $n_\pi \pi \to p\bar p$ with an 
average of $n_\pi\simeq$~5-6 pions in the incoming channel. 
In the following we will show that the inclusion of these reactions, 
treated within a thermal rate equation, is indeed capable of 
regenerating a significant number of antiprotons~\cite{RS01}. 
The notion of detailed balance alone, however, is not enough. 
A second crucial 
ingredient~\cite{RS01} is the over-saturation of pion phase-space, 
encoded in nonzero pion-chemical potentials, 
which enhances the equilibrium abundance of antiprotons substantially 
(see also ref.~\cite{Cass02} for a recent transport approach to the 
problem).

\section{Thermal $\bar p$-Production}
\label{sec_rateq}
\subsection{Rate Equation and Chemical Potentials}
In thermal equilibrium, the (net) rate per unit 4-volume for 
producing antiprotons via $n_\pi$-pion fusion and its
back-reaction, $n_\pi \pi \leftrightarrow p\bar p$, is given
in terms of the corresponding invariant matrix element 
${\cal M}_{n_\pi}$ as (see, \eg, ref.~\cite{KMR86})
\bea
{\cal R}_{\bar p}^{th} = \int d^3\tilde k_p d^3\tilde k_{\bar p}
\prod\limits_{i=1}^{n_\pi} d^3\tilde k_{i}
~\delta^{(4)}(K_{tot})~|{\cal M}_{n_\pi}|^2~
\{ z_p z_{\bar p} \ {\rm e}^{-\frac{E_p+E_{\bar p}}{T}} -
z_\pi^{n_\pi} {\rm e}^{-\sum\limits_{i=1}^{n_\pi}\frac{\omega_i}{T}}\} \ . 
\label{rate}
\eea
The thermal factors are in Boltzmann approximation, which
allows to factorize the fugacity factors $z_i=\exp(\mu_i/T)$. 
The rate equation can then be written in the compact form
\beq
{\cal R}_{\bar p}^{th} = \frac{d\rho_{\bar p}}{dt} \simeq 
\frac{\rho_{\bar p}}{\tau_{\bar p}^{chem}} 
\left( \frac{\langle z_\pi^{n_\pi} \rangle}{z_p z_{\bar p}} -1 \right) \ ,
\label{rate2}
\eeq
which represents a first
order differential equation in the antiproton density $\rho_{\bar p}$. 
The relaxation time $\tau_{\bar p}^{chem}$ is given in  
eq.~(\ref{tauchem}); the use of the total baryon density $\rho_B$ therein 
(rather than just the proton density $\rho_p$, which is much smaller) 
is based on the assumption that the annihilation cross section of 
antiprotons on other baryons is similar to the one on protons. 
Eq.~(\ref{rate2}) furthermore presumes that $\bar p$ annihilation on 
excited resonances yields (on average) as many additional pions 
as the resonance itself decays into (\eg, $\bar p$ annihilation on 
a $\Delta(1232)$ gives one pion more than on a nucleon). This leads 
to  a cancellation of additional fugacity factors multiplying 
$\langle z_\pi^{n_\pi} \rangle$ and $z_p$ in eq.~(\ref{rate2}). 
The use of the {\em total} annihilation cross section 
in $\tau_{\bar p}^{chem}$ implies
a sum over all possible $n_\pi$-pion final states ($s\bar s$ production
is suppressed by the OZI rule). Again, this automatically incorporates 
mesonic resonance states such as $\rho$ and $\omega$ to the extent
that they contribute with a pion-fugacity factor according to their
pion final states. In a heavy-ion environment this is obviously satisfied
for $\rho$ mesons, but to a good approximation also for $\omega$ (and even
$\eta$) mesons, cf.~footnote 1.

Under SPS conditions, where the relative amount of antiprotons in the 
system is small, baryon and pion densities are not significantly 
altered by the reactions under consideration. The rate equation can 
thus be considered with the anitproton density as its only variable.
In equilibrium one has ${\cal R}_{\bar p}^{th}=0$, so that the 
antiproton fugacity can be determined as
\beq
z_{\bar p}^{eq}= z_p \ \langle z_\pi^{n_\pi} \rangle \ . 
\label{zpbar}
\eeq
Here, as in eq.~(\ref{rate2}), the averaging of $z_\pi^{n_\pi}$ 
is over the initial (or final) pion numbers. 
Since the pion fugacities enter with large powers, a reasonable 
estimate of $z_{\bar p}$ requires a reliable description of the
final state multiplicities in $p\bar p$ annihilation, which we 
will turn to in the following section.

\subsection{Properties of $\bar p p$ Annihilation}
Fig.~\ref{fig_dist} shows the experimental results for pion
multiplicities in $p\bar p$ annihilation at rest (see, 
\eg, the review \cite{Dov92} and references therein). 
The distribution can be reasonably well reproduced by a Gaussian
$P(n_\pi)$ with an average $\langle n_\pi\rangle=5.06$ and a 
width $\delta n_\pi =0.9$. 
\bfg[!ht]
\begin{minipage}[t]{6.8cm}
\vspace{0.27cm}
\epsfig{file=Pnpi-fit.eps,width=6.75cm}
\vspace{-0.69cm}
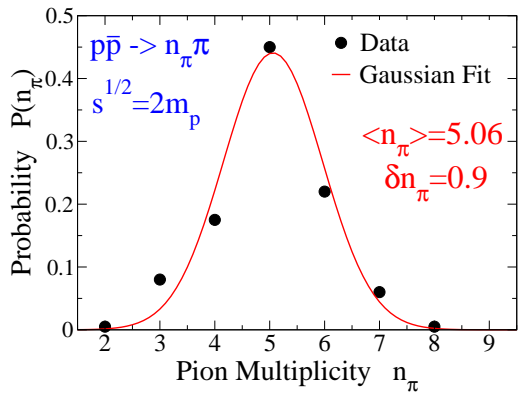
\caption{Pion-multiplicity distribution in $p\bar p$ annihilation 
at rest, together with a Gaussian fit using the indicated mean and width.}
\vspace{-0.2cm}
\label{fig_dist}
\end{minipage}\hfill
\begin{minipage}[t]{6.8cm}
\vspace{0.2cm}
\epsfig{file=ppb_npi.eps,width=6.8cm}
\vspace{-0.69cm}
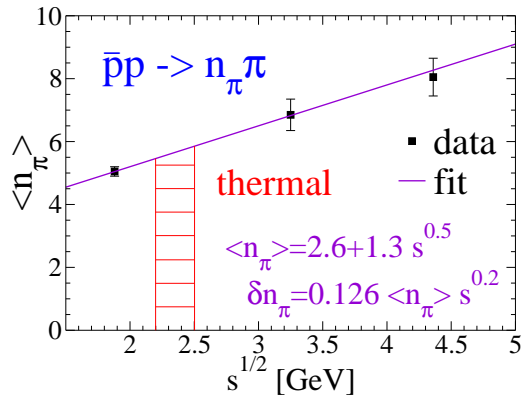
\caption{Dependence of average pion-multiplicity in $p\bar p$ annihilation
on CMS energy. The fit and corresponding parameterizations are also indicated.}
\vspace{-0.2cm}
\label{fig_npis}
\end{minipage}
\efg

In a thermal heat bath, the annihilation
processes occur, on average, at somewhat larger CMS energies,
$\sqrt{s}\simeq 2 E_p^{th}$.
Experimental data on the $s$-dependence
of $\langle n_\pi\rangle$ are displayed in Fig.~\ref{fig_npis}. A linear
increase accurately describes $\langle n_\pi\rangle(s) $ in the relevant
(thermal) energy range. At the same
time, also the width of the Gaussian moderately increases.
Put together, the averaged multi-pion fugacity at given temperature
takes the form
\beq
\langle z_\pi^{n_\pi} \rangle(T) = \sum\limits_{n_\pi=2}^{n_\pi^{max}}
P(n_\pi) \ {\rm e}^{n_\pi\mu_\pi/T} \ ,
\label{zpi_av}
\eeq
where for any practical purposes $n_\pi^{max}=9$.

For the treatment of off-equilibrium effects (cf.~Sect.~\ref{offeq}) one
also needs to account for the energy-dependence of the total annihilation 
cross section, which figures into the equilibration time, 
eq.~(\ref{tauchem}). Its decrease with increasing $s$
can be fitted by the empirical formula 
$\sigma_{p\bar p}^{ann}(s)=(40p_{lab}^{-0.5} +24~p_{lab}^{-1.1})$~mb.

\subsection{$\bar p$ Thermodynamics at SPS Energies}
Inserting the pion-fugacity factor, $\langle z_\pi^{n_\pi} \rangle(T)$
from eq.~(\ref{zpi_av}) into eq.~(\ref{zpbar}), and using the
temperature-dependent pion- and nucleon-chemical potentials as
displayed in Fig.~\ref{fig_chemevo}, results in an equilibrium
$\bar p/p$ ratio for central Pb(158~AGeV)+Pb collisions as 
represented by the solid lines in Fig.~\ref{fig_ratioSPS}. 
\bfg[!ht]
\begin{minipage}[t]{6.8cm}
\vspace{-0.3cm}
\epsfig{file=pbpratio-SpS.eps,width=6.8cm}  
\vspace{-0.49cm}
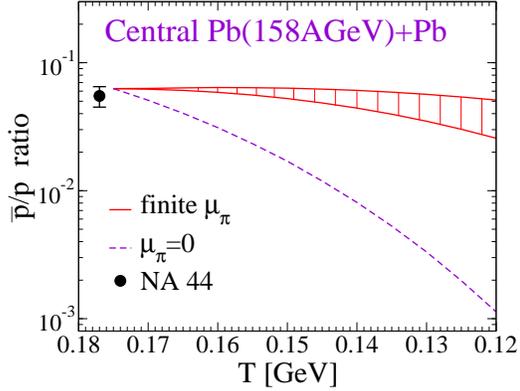
\caption{Equilibrium $\bar p/p$ ratio at SPS energies with zero (dashed line)
and finite (solid lines) pion-chemical potentials (the hatched region 
corresponds to $\mu_\pi^{therm}$=65-80~MeV).}
\vspace{-0.2cm}
\label{fig_ratioSPS}
\end{minipage}\hfill
\begin{minipage}[t]{6.8cm}
\vspace{-0.42cm}
\epsfig{file=tauchem-SpS.eps,width=6.6cm}
\vspace{-0.0cm}
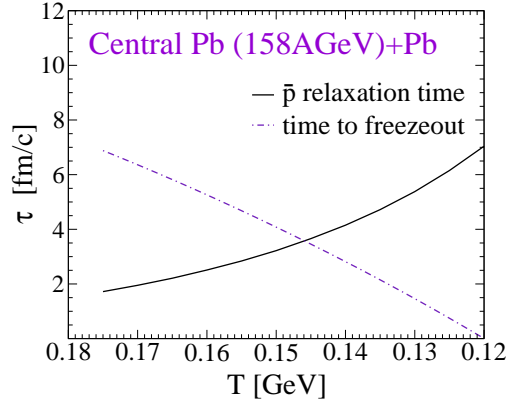
\caption{Chemical relaxation time for antiprotons (solid line),
eq.~(\protect\ref{tauchem}), and remaining fireball lifetime until 
thermal freezeout (dashed-dotted line) at SPS energies~\protect\cite{RW99}.}
\vspace{-0.2cm}
\label{fig_tauSpS}
\end{minipage}
\efg
The enclosed band approximately reflects the uncertainties in 
determining the pion-chemical potential, associated with the population 
of higher resonance states (especially those with relatively long lifetime)
in the late stages of the hadronic evolution.
Irrespective, the strong enhancement over the equilibrium value
without the inclusion of finite $\mu_\pi$'s (dashed line in
Fig.~\ref{fig_ratioSPS}) is obvious.

In Fig.~\ref{fig_tauSpS}, the chemical relaxation time is compared 
to the fireball lifetime, indicating that the antiproton abundance  
might be driven out of equilibrium before thermal freezeout. 
This will be quantitatively assessed in the following 
section.

\subsection{Off-Equilibrium Evolution}
\label{offeq}
For an explicit solution of the rate equation describing the 
evolution of the antiproton number in an expanding hadron gas
we rely on a hierarchy of time scales 
$\tau^{therm} < \tau_{\bar p}^{chem} \ll \tau_\pi^{chem}$. This  
implies an equilibrium $\bar p$ density 
\beq
\rho_{\bar p}^{eq} = \langle z_\pi^{n_\pi} \rangle
\int \frac{d^3k}{(2\pi)^3} \ \exp[-(E_{\bar p}-\mu_{\bar p})/T] \ .  
\eeq
The finite system size in a heavy-ion collision 
requires the inclusion of volume expansion effects~\cite{KMR86}. Using 
$ \ dN_{\bar p}= {\cal R}_{\bar p}^{th} \ V_{FB}(t) \ dt \ $ 
where $N_{\bar p}= \rho_{\bar p} V_{FB}$ is the antiproton number and 
$V_{FB}(t)$ the expanding fireball volume~\cite{RW99}, one obtains
\beq
\frac{d\rho_{\bar p}}{dt}= -\rho_{\bar p} 
\left( \frac{1}{\tau_{\bar p}(t)} + \frac{1}{V_{FB}(t)} 
 \ \frac{dV_{FB}}{dt} \right) + 
\rho_{\bar p}^{eq}(t)\frac{1}{\tau_{\bar p}(t)}
\equiv -\rho_{\bar p}(t) L(t) + G(t) 
\label{diffeq}
\eeq 
with gain and loss terms $G(t)$ and $L(t)$, respectively. 
Eq.~(\ref{diffeq}) is an ordinary first order differential equation which
has the solution
\beq
\rho_{\bar p}(t) =\varphi(t) \left( \rho_{\bar p}(t_0) + 
\int\limits_{t_0}^t dt' \ G(t')/ \varphi(t') \right) \ , \quad 
\varphi=\exp[-\int\limits_{t_0}^t dt' \ L(t')] \ . 
\label{solu}
\eeq
\bfg[!ht]
\epsfig{file=pbpevo-SpS.eps,width=5.3cm,angle=-90}
\epsfig{file=pbpevoXS-SpS.eps,width=5.3cm,angle=-90}
\vspace{-0.2cm}
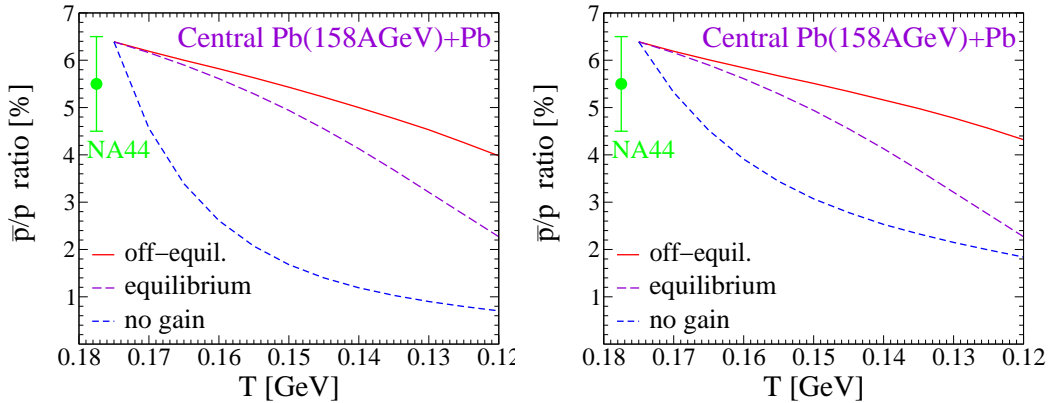
\caption{Solid lines: evolution of the $\bar p/p$ ratio according to the 
solution, eq.~(\protect\ref{solu}), of the rate eq.~(\protect\ref{diffeq})
for an expanding fireball using the free (left panel) and a 50\%
reduced (right panel) $\sigma_{\bar p p}^{ann}$; long-dashed 
lines: equilibrium ratio; short-dashed lines: solution when switching off 
the gain term $G(t)$.}   
\vspace{-0.5cm}
\label{fig_offeq}
\efg
As shown in Fig.~\ref{fig_offeq}, the
off-equilibrium results confirm the importance of both the
pion-fugacities as well as the backward reaction of multi-pion
fusion in maintaining an antiproton abundance close to the observed
value. The effective freezeout temperature $T_{\bar p}^{chem}$ 
turns out to be close to 130~MeV (slightly higher if one assumes a, 
\eg, 50\% smaller in-medium annihilation cross section).

\section{Summary}
\label{sec_sum}
We have shown that antiproton production in central 
heavy-ion collisions at SPS energies can be reconciled with the observed
value corresponding to standard chemical freezeout: annihilation 
losses in the subsequent hadron gas phase are compensated by the 
back-reaction of multipion fusion, reinforced by the build-up of 
large pion-chemical potentials. 
We have corroborated our earlier findings for the equilibrium case 
by an explicit solution of the underlying rate equation. 
An interesting test of the regeneration mechanism could be provided 
by the recently suggested balance function technique~\cite{BDP01}, which 
has significant sensitivity to the time of (pair) production especially 
for heavy particles (such as nucleons). 
The off-equilibrium framework is more involved under RHIC 
conditions, as $\rho_B$ is of the same order as 
$\rho_{\bar B}$, so that feedback effects on the baryon densities cannot
be neglected. This will be addressed in future work, as it promises 
valuable insights into hadro-production at collider energies.

\vspace{0.5cm}

\noindent
{\bf ACKNOWLEDGMENT} \\
One of us (RR) would like to thank the organizers for the invitation
and a very pleasant and informative meeting.
This work has been supported by the US Department of Energy under
grant number DE-FG0288ER40388.

\end{document}